\documentclass[prl,showpacs,superscriptaddress,twocolumn]{revtex4-1}
\usepackage[pdftex]{graphicx}  
\usepackage{graphicx, xcolor}
\usepackage{dcolumn}
\usepackage{bm}
\usepackage{amsmath,amsthm,amssymb}
\usepackage{color}
\usepackage{verbatim}
\usepackage{natbib}

\definecolor{darkGreen}{RGB}{0,110,0}
\definecolor{darkBlue}{RGB}{0,0,130}
\usepackage[colorlinks,citecolor=darkGreen,linkcolor=darkBlue,urlcolor=blue,hyperindex]{hyperref}

\def\be{\begin{equation}}
\def\ee{\end{equation}}

\newcommand{\ket}[1]{\left| #1 \right\rangle}

\begin{document}
\title{Stark many-body localization}

\author{M. Schulz}
\affiliation{SUPA, School of Physics and Astronomy, University of St Andrews, North Haugh, St Andrews, Fife KY16 9SS, United Kingdom}
\affiliation{Max Planck Institute for the Physics of Complex Systems, N{\"o}thnitzer Str.\ 38, 01187 Dresden, Germany}
\author{C.A. Hooley}
\affiliation{SUPA, School of Physics and Astronomy, University of St Andrews, North Haugh, St Andrews, Fife KY16 9SS, United Kingdom}
\author{R. Moessner}
\affiliation{Max Planck Institute for the Physics of Complex Systems, N{\"o}thnitzer Str.\ 38, 01187 Dresden, Germany}
\author{F. Pollmann}
\affiliation{Physics Department, Technical University of Munich, James-Franck-Str.\ 1, 85748 Garching, Germany}

\date{Monday 18th March 2019}

\begin{abstract}
\noindent
We consider spinless fermions on a finite one-dimensional lattice, interacting via nearest-neighbor repulsion and subject to a strong electric field.  In the non-interacting case, due to Wannier-Stark localization, the single-particle wave functions are exponentially localized even though the model has no quenched disorder.  We show that this system remains localized in the presence of interactions and exhibits physics analogous to models of conventional many-body localization (MBL).  In particular, the entanglement entropy grows logarithmically with time after a quench, albeit with a slightly different functional form from the MBL case, and the level statistics of the many-body energy spectrum are Poissonian.  We moreover predict that a quench experiment starting from a charge-density wave state would show results similar to those of Schreiber {\it et al.\/} [{\it Science\/} {\bf 349}, 842 (2015)].
\end{abstract}

\maketitle

\paragraph{Introduction.}
In the latter half of the twentieth century, pioneering work by Anderson \cite{Anderson1958} and collaborators \cite{Abrahams1979} established that the eigenfunctions of a quantum particle moving in a disordered potential landscape in one or two dimensions are exponentially localized.  In the beginning of the twenty-first century, it was shown \cite{Basko2006,Gornyi2005,Imbrie2016} that this localization can persist even for finite densities of fermions with repulsive interparticle interactions, a phenomenon now termed many-body localization (MBL) \cite{Nandkishore2015}.

Subsequent theoretical work has explored the disorder-driven transition from ergodic to MBL behavior in a variety of mostly one-dimensional models.  Key results include the `l-bit' picture of the MBL regime \cite{Huse2014a}, the logarithmic growth of entanglement entropy following a quench into it \cite{Znidaric2008,Bardarson2012,Serbyn2013a}, and subdiffusive transport of spin and energy in the approach to the MBL transition from the ergodic side \cite{Vosk2015,Agarwal2015,Znidaric2016a,Bar-Lev2015}.  Significant progress has also been made in the experimental study of MBL in one- and two-dimensional systems of cold atoms \cite{Schreiber2015,Bordia2017,Choi2016}.

It is natural to ask whether disorder is at all an essential ingredient for observing MBL phenomenology, or whether it is possible for thermalization to fail for reasons due to interactions alone.  This can happen when the system is integrable~\cite{Rigol2007,Kinoshita2006}, but such cases constitute isolated points in parameter space, with arbitrarily small generic variations in the Hamiltonian parameters restoring ergodicity.  An important question is then whether there can be {\it robust\/} non-ergodic phases in models without quenched disorder.  There have so far been several proposals for this, ranging from models inspired by classical glassy physics \cite{VanHorssen2015,Hickey2016a} to variants of the Bose Hubbard model \cite{DeRoeck2014,Gornyi2005} and models with mixed species~\cite{Smith2017}; however, there are also counterarguments that the localization in some of these systems is a long-lived intermediate-time phenomenon which eventually yields to thermalization at very long times \cite{Yao2016,Bols2018}.

Here, we approach the question of MBL without disorder from a different angle, by introducing interactions into a single-particle model that exhibits Wannier-Stark localization \cite{Wannier1962}.  It is instructive to compare the resulting physics, which we call Stark many-body localization, with the well known many-body disorder localized (MBL) case.  The comparison is interesting because in both cases the single-particle eigenfunctions are exponentially localized in space, and thus one might expect at least some MBL signatures also to be present in the disorder-free Stark-MBL case.

\begin{figure}
	\includegraphics[width=\linewidth]{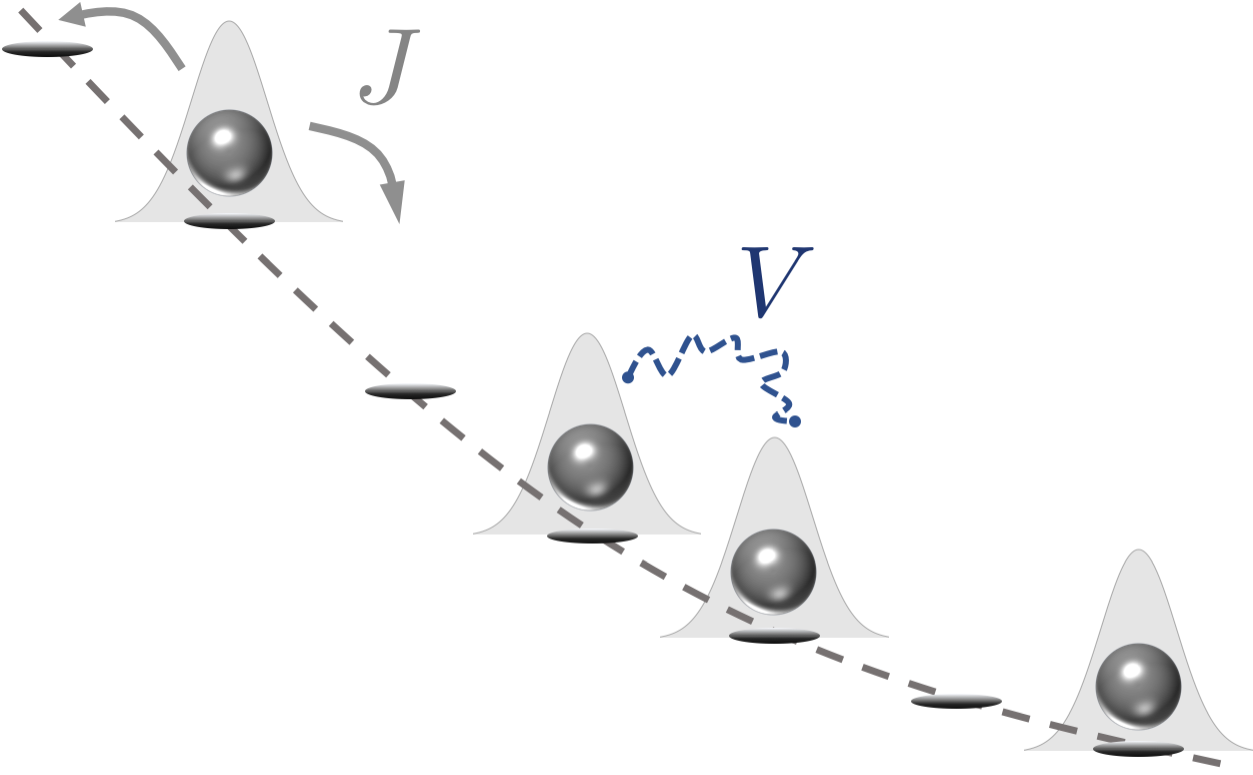}
	\caption{Schematic illustration of our model. On a \textit{finite} lattice in one dimension, the single particle states of spinless fermions, which can delocalize via hopping $J$, are exponentially localized via a strong electric field. For a purely uniform field (i.e.\ a linear electric potential), this is usually referred to as Wannier-Stark localization \cite{Wannier1962}. When the particles interact via nearest-neighbor interactions $V$, they share many properties with the well studied MBL phenomenology.}
	\label{fig:system}
\end{figure}

Specifically, we study the properties of spinless fermions on a finite one-dimensional lattice.  The fermions interact via nearest-neighbor repulsion, and we subject them to a strong electric field:\ either a strictly spatially uniform one, for which the non-interacting many-body spectrum contains many exact degeneracies, or a field with a slight spatial gradient, for which these degeneracies are lifted. The setup is depicted in Fig.~\ref{fig:system}.

\begin{figure}
	\includegraphics[width=\linewidth]{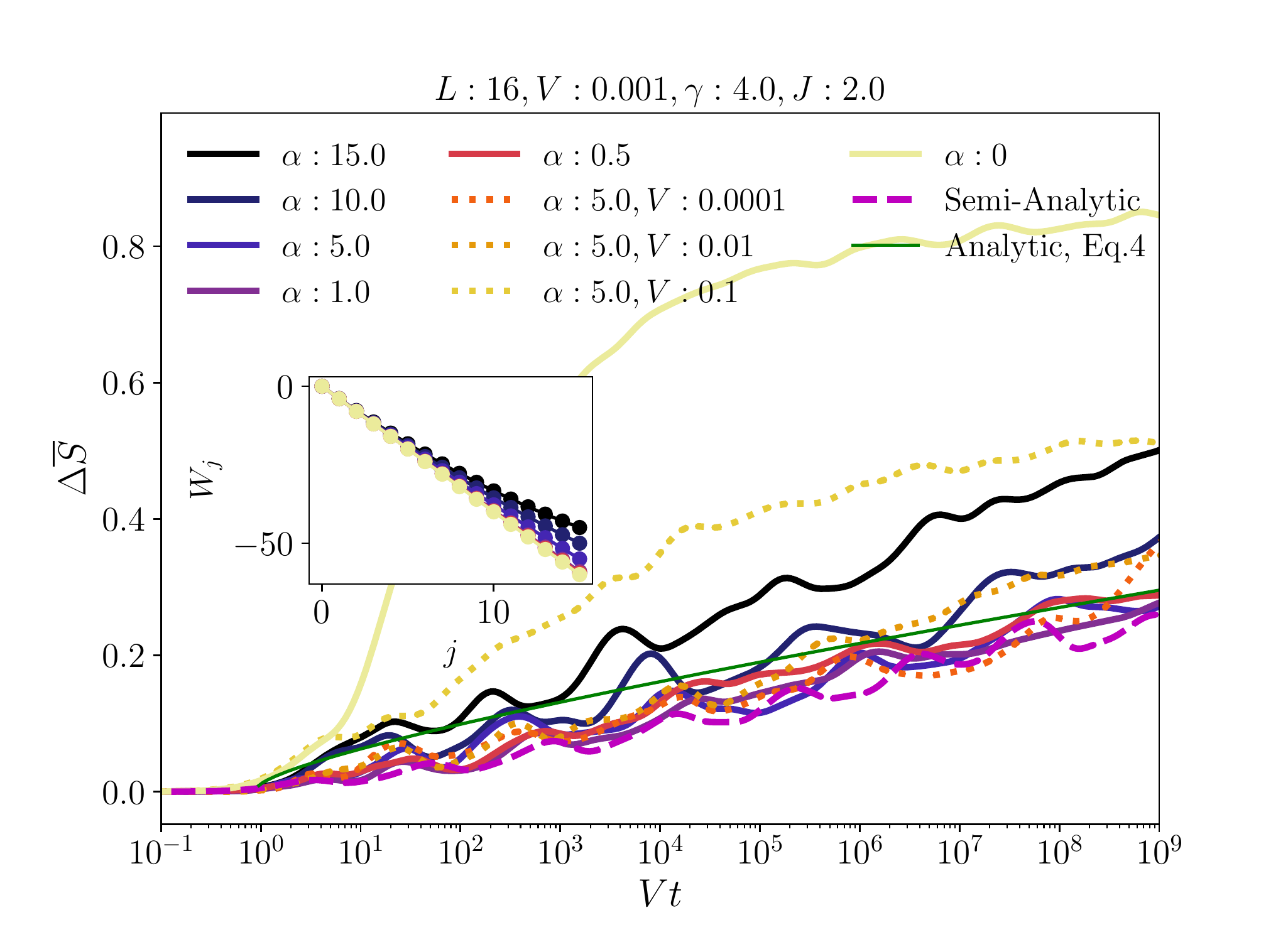}
	\caption{Difference $\Delta S$ between the bipartite entanglement in the interacting and non-interacting cases. The parameter $\alpha$ denotes the cumulative effect of field non-uniformity on the potential at the end site. For sufficiently small $\alpha$ and suitably large field $\gamma$ there is good qualitative agreement with our semi-analytic calculation (magenta dashed line) and full analytic calculation (green line and Eq.~\eqref{equ:entgrowth}). For larger $\alpha$, the entanglement growth becomes stronger than predicted due to the progressive delocalization of some of the single-particle orbitals as the position-dependent field gets weaker at the right-hand end of the chain (see right of inset). For the purely uniform field, there is an initial steep rise which we attribute to additional degeneracies of the many-body spectrum. The numerical curves have been smoothed by convolution with a Gaussian, $w(n) = e^{-(n/\sigma)^2/2}$, with $\sigma = 4$. Inset: Potential used in the respective main curves.}
	\label{fig:log_growth}
\end{figure}

\begin{figure}
	\includegraphics[width=\linewidth]{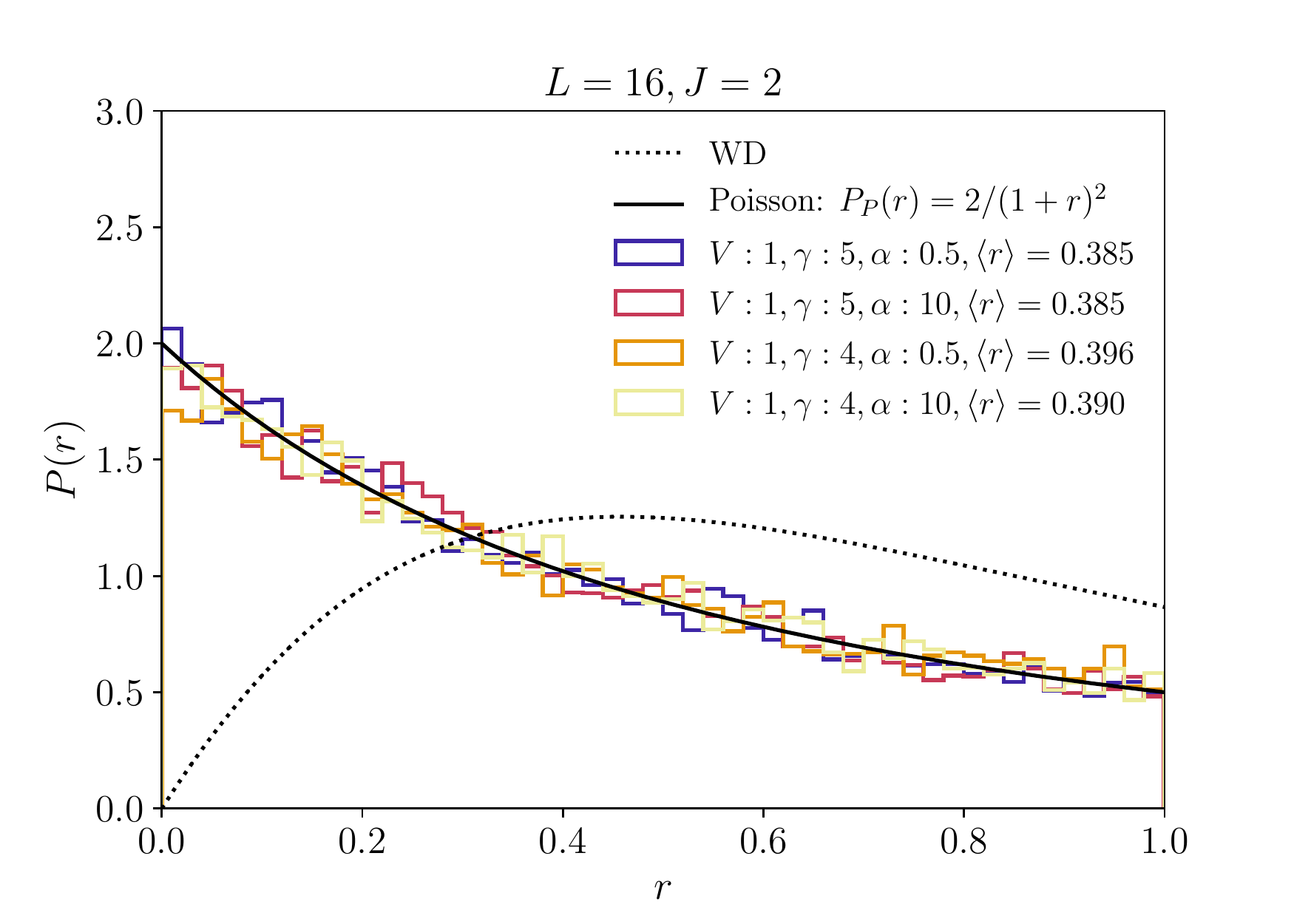}
	\caption{The many-body level statistics for the case of nonzero $\alpha$ in the half-filled sector. For all displayed values of $\gamma$ and $\alpha$ the probability distribution of the gap-ratio parameter, $r_n = \min(\delta_n/\delta_{n+1}, \delta_{n+1}/\delta_n)$ (where $\delta_n$ is the gap between the $n^{\text{th}}$ and $(n-1)^{\text{th}}$ energy eigenvalues), agrees with the prediction for Poisson level statistics expected for integrable or localized models. The predicted average gap-ratio parameter in this case is $\langle r \rangle_{P} = 0.3863$. For comparison we have included the prediction for Wigner-Dyson (WD) statistics.}
	\label{fig:level_stats}
\end{figure}

For fields with slight spatial gradients, we find that Stark-MBL is robust and shares many similarities with MBL.  Specifically, the entanglement entropy grows logarithmically with time after a quench (Fig.~\ref{fig:log_growth}), and the many-body eigenenergies of the Stark-MBL problem generically show Poissonian level statistics (Fig.~\ref{fig:level_stats}).  We moreover predict the results in our Stark-MBL system of a quench experiment starting from a charge-density wave state of the type conducted in~\cite{Schreiber2015}, and we obtain graphs similar to those seen in the MBL case. Interestingly, for a strictly uniform field~\cite{udssr}, entropy and level statistics deviate from MBL phenomenology, which we attribute to exact degeneracies in the non-interacting spectrum.

\paragraph{Model.} The Hamiltonian of our model is
\begin{multline}
\label{equ:hamiltonian}
\hat{H} =  \frac{J}{2} \,\sum_{j=0}^{L-2} \left( c^\dagger_j c^{\phantom{\dagger}}_{j+1} + \text{h.c.} \right)  + \sum_{j=0}^{L-1} W_j \left(n_j -\frac{1}{2}\right) \\ + V \sum_{j=0}^{L-2} \left(n_j -\frac{1}{2}\right) \left(n_{j+1} -\frac{1}{2}\right)  . 
\end{multline}
Here the operator $c^{\dagger}_{j}$ creates a fermion on lattice site $j$, and the associated number operator $n_j = c^{\dagger}_{j} c_{j}$. $J/2$ is the hopping matrix element between neighboring sites, $V$ is the strength of the nearest-neighbor repulsion, and $L$ is the number of sites of the lattice, the boundary conditions of which we take to be open.

$W_j$ is the on-site potential energy due to the applied electric field.  For a uniform field $\gamma$, it takes the form $W_j = -\gamma j$.  For a non-uniform field, it acquires some curvature, $W_j = - \gamma j + \alpha j^2/(L-1)^2$.   This means that the potential has the same value and slope as the linear one at $j=0$, but differs at the other end of the chain by $\alpha$ from the purely linear case (see the inset of Fig.~\ref{fig:log_growth}).

\paragraph{Entanglement entropy growth.}  A quantity that is often used to diagnose and characterize MBL is the entanglement entropy.  To calculate this, a spatial bipartition into two halves is made by cutting a particular bond.  The reduced density matrix of the subsystem, $\rho_{\rm red}$, is calculated, and the entropy determined via $S=-\text{tr}\left(\rho_{\text{red}}~\text{log}(\rho_{\text{red}})\right)$.  Serbyn {\it et al.\/} \cite{Serbyn2013a} argued, on the basis of the general physical picture of `l-bits' in the MBL regime, that at long times the entropy should grow logarithmically with time, i.e.\ that $S(t) \sim S_0 \ln (Vt)$.

This argument uses the fact that the density profiles of the individual l-bits are exponentially localized.  We now show that the long-time entanglement growth in the Stark-MBL system behaves similarly, though with some modification reflecting the fact that the tails of Wannier-Stark-localized wave functions are not pure exponentials.

The functional form of these tails may be obtained from a semiclassical argument \cite{Hooley2004} in which the wave function is approximated by a WKB-like expression
$\psi(x) \sim \exp \left( i \int^x k(x') dx' \right),$
with $k(x)$ obtained from the condition of local energy conservation, $J \cos \left( ka \right) + V(x) = E$, where $a$ is the lattice spacing.
Our potential is to a good approximation $V(x) = -\gamma x$, and hence
$ k(x) = \frac{1}{a} \arccos \left( \frac{\gamma x + E}{J} \right)$.
Taking $x$ to be large and positive, the exponent in the WKB wave function is thus given by $- \frac{1}{a} \int^x \ln \left( \frac{\gamma x'}{J} \right) dx' \approx
- \frac{x}{a} \ln \left( \frac{\gamma x}{J} \right),$
where we have dropped a linear-in-$x$ term that does not have a logarithm.  Hence the form of the tail of the wave function is
\begin{equation}
\psi(x) \sim \exp \left[ - \frac{x}{a} \ln \left( \frac{\gamma x}{J} \right) \right]. \label{tail}
\end{equation}

We now use (\ref{tail}) to derive the form of the entanglement entropy $S(t)$ at long times, in analogy to Ref.~\cite{Serbyn2013a}.  The matrix element of the nearest-neighbor repulsion between a pair of localized particles at distance $x$ is approximately $V \exp \left[ - \frac{x}{a} \ln \left( \frac{\gamma x}{J} \right) \right]$, corresponding to a dephasing time of
\begin{equation}
t_{\rm deph} \approx V^{-1} \exp \left( \frac{x}{a} \ln \left( \frac{\gamma x}{J} \right) \right). 
\label{dephtime}
\end{equation}
This is the timescale on which the entanglement entropy due to this pair will rise from zero to its maximum value.

Now consider the total entanglement entropy of the subsystem.  This can be thought of as an entanglement front moving through the system from the cut as $t$ increases.  At time $t$, this front will have moved a distance $x(t)$, which is obtained by solving (\ref{dephtime}) for $x$. The entanglement entropy will thus be $S(t) = S_{\rm max} \frac{x(t)}{L},$ where $L$ is the length of the subsystem to the left of the cut. Solving (\ref{dephtime}) for $x(t)$, we obtain as entanglement entropy:
\begin{equation}
\label{equ:entgrowth}
S(t) = \frac{S_{\rm max}}{L} \frac{a \ln(Vt)}{p \left( \frac{a \gamma \ln (Vt)}{J} \right)},
\end{equation}
where $p(u)$ is the solution of the equation $p\,e^p = u$.

We supplement this analytic approach by computing a `semi-analytic' form for the entanglement entropy.  For strong fields, $\gamma \gg J$, we may assume that the single-particle eigenfunctions of the Stark-localized states are restricted to three sites.  We can thus write any initial state with fixed particle number $M$ in fixed positions as
\begin{equation}
\label{equ:starkproj}
\ket{\psi(t=0)} = \sum_{\{Q\}} f_Q \beta^\dagger_{Q(j_{M})} \beta^\dagger_{Q(j_{M-1})} \ldots \beta^\dagger_{Q(j_{2})} \beta^\dagger_{Q(j_{1})} \ket{0},
\end{equation}
where $\beta^\dagger_j$ creates a fermion in the single-particle Stark-localized state centered at position $j$.  Here $Q$ is an operation that, for each original site label $j_i$, either moves it one place to the left, moves it one place to the right, or leaves it alone. The factor $f_Q$ is given approximately by
\begin{equation}
f_Q \approx \left( \frac{J}{\gamma} \right)^{\lambda_Q + \rho_Q}, \label{stateweight}
\end{equation}
where $\lambda_Q$ counts the number of original site labels that are shifted to the left by $Q$, and $\rho_Q$ the number that are shifted to the right.

The energy $E_Q$ of every state-component $Q$  is composed of two parts:\ the electric potential energy, and the repulsive interaction energy of each pair of its constituent particles:
\begin{equation}
\label{equ:mbenergy}
E_Q \approx E_0 + \gamma \left( \rho_Q - \lambda_Q \right) + \frac{V}{2} \sum_k \sum_{p \ne k} \left( \frac{J}{\gamma} \right)^{2 \left\vert Q(j_k) - Q(j_p) \right\vert -2},
\end{equation}
where $E_0$ is the energy of some reference state and the sums over $k$ and $p$ run from $1$ to $M$.

Combining \eqref{equ:starkproj} and \eqref{equ:mbenergy}, we obtain the initial state time-evolved to arbitrary times:
 \begin{equation}
 \label{equ:icarbtime}
\ket{\psi(t)} = \sum_Q f_Q e^{-i E_Q t} \beta^\dagger_{Q(j_{M})} \beta^\dagger_{Q(j_{M-1})} \ldots \beta^\dagger_{Q(j_{2})} \beta^\dagger_{Q(j_{1})} \ket{0}.
\end{equation}
This allows us to calculate the time-evolved reduced density matrix of any subsystem, from which we obtain $S(t)$ via diagonalization.
The results can be found in Fig.~\ref{fig:log_growth} and further details on the above calculation are found in~\cite{supp}.
\paragraph{Numerical simulations.}
We now compare the above results against numerical simulations of the model \eqref{equ:hamiltonian}. Hopping is set to $J = 2$ and we consider an $L = 16$ chain at half-filling. The field on one end has strength $\gamma$ and decreases uniformly, so that the value of the potential on the final site of the chain differs by an amount $\alpha$ from its value in the purely linear case. We use all possible initial product states that do not have a particle on the two sites directly adjacent to the cut, which we choose to be at the middle bond. For $L = 16$, this amounts to 1225 states.
Using exact diagonalization, we compute the time-evolution for a given initial state, and obtain the entanglement entropy $S(t)$ for an equal bipartition of the lattice. We then average over all of the abovementioned initial states.

Our results for $\gamma = 4$ and various values of $\alpha$ are shown in Fig.~\ref{fig:log_growth}. In order to show the effect of interactions, we display $\Delta \overline{S}(t) \equiv \overline{S}(t)-\overline{S}_0(t)$, where $S_0(t)$ is the entropy in the non-interacting case, and the bar denotes an average over the initial conditions as discussed above.  Consistent with previous findings, the entanglement growth due to many-body dephasing effects commences at times $Vt \sim 1$. The curves scale as a function of $Vt$, until the interactions become strong enough to alter the eigenstates significantly.

For sufficiently nonuniform field, this growth continues logarithmically up to a maximum value , which is finite for a finite system size. This growth agrees qualitatively with our semi-analytic calculation presented above, confirming our intuition that the physics of this system is well described by \eqref{equ:starkproj} and \eqref{equ:mbenergy}.  We have also compared the pure analytic calculation \eqref{equ:entgrowth} against the numerics. Here $S_\text{max}$ is taken to be as the diagonal entropy $S_\text{diag}$ resulting from the reduced density matrix obtained by \eqref{equ:icarbtime}. $S_\text{diag}$ denotes the maximum achievable entropy for a given initial state, assuming that the interactions do not significantly change the form of the eigenstates \cite{Polkovnikov2011a,Serbyn2013a}.

In the case of a strictly linear potential, there is instead a steep rise of the entanglement entropy followed by slow growth. We attribute this to the many exact degeneracies in the non-interacting version of the problem, which arise from the fact that all two-particle states with the same center of mass are degenerate, and can therefore hybridize strongly.

The logarithmic entanglement growth given by \eqref{equ:entgrowth} can thus fail for two different reasons:\ too little non-uniformity, or too small a local field.  The former allows delocalization of the two-particle states into a broad band, with a correspondingly short dephasing time, leading to steep initial growth of $\Delta \overline{S}(t)$.  On the other hand, the latter causes the localization to weaken in parts of the chain, in turn making the accumulation of entanglement in that part stronger.  The former case is shown in Fig.~\ref{fig:log_growth} for $\alpha = 0$, while the latter can be observed in the same panel for $\alpha = 15$. However, in between them there is a large range of curvature values for which an MBL-like entanglement growth is observed.

\paragraph{Many-body level statistics.} 
A powerful and basis-independent diagnostic to determine whether a model is localized is its spectral statistics \cite{Oganesyan2007, Luitz2015}. Fig.~\ref{fig:level_stats} shows our exact diagonalization results for the gap-ratio parameter in a chain of length $L=16$ with open boundary conditions, for a range of field strengths $\gamma$ and gradients $\alpha$ in the half-filled sector. The gap-ratio parameter is defined as $r_n = \min(\delta_n/\delta_{n+1}, \delta_{n+1}/\delta_n)$, where $\delta_n$ is the gap between the $n^{\text{th}}$ and $(n-1)^{\text{th}}$ energy eigenvalues. Plotting a histogram of the gap-ratio parameter eliminates a dependence on the density of states, such that we can use the whole spectrum as long as there is no mobility edge present. We have verified this and the results are qualitatively the same for all ranges and energy densities used.

The probability distribution for Poisson statistics reads $P_P(r) = 2/(1+r)^2$, which is in excellent agreement with our results over most of our parameter range. In the case of very small or zero field gradient $\alpha$ however, we find results that are consistent neither with Poissonian nor with Wigner-Dyson (WD) statistics. We attribute this to the many exact degeneracies in the non-interacting problem creating a disproportionally large weight for small $r_n$ in the many body spectrum. Interestingly we may further observe a cross-over to WD level statistics for small uniform field strength $\gamma$. Results for both of these cases can be found in the Supplemental Material~\cite{supp}.

\paragraph{Quench from charge-density wave.}  
While the entanglement and spectral statistics are experimentally difficult to extract, there are other indicators of localization.
One of the simplest setups consists of monitoring the relaxation of an initial charge-density wave (CDW) order, in which all even sites are occupied \cite{Schreiber2015}. An associated observable is the imbalance $\mathcal{I}$ between the occupation on odd and even sites, $N_{\text{o}}$ and $N_{\text{e}}$ respectively,
\begin{equation}
	\label{equ:imbalance}
	\mathcal{I} = \frac{N_{\text{e}}-N_{\text{o}}}{N_{\text{e}}+N_{\text{o}}}.
\end{equation}
In the ergodic, thermalizing case it should quickly decay to zero, which it does in the case $\gamma=0$ when there is no Wannier-Stark localization.
For a many-body localized system, on the other hand, the value of the imbalance should remain non-zero up to infinite times.

\begin{figure}
	\includegraphics[width=\linewidth]{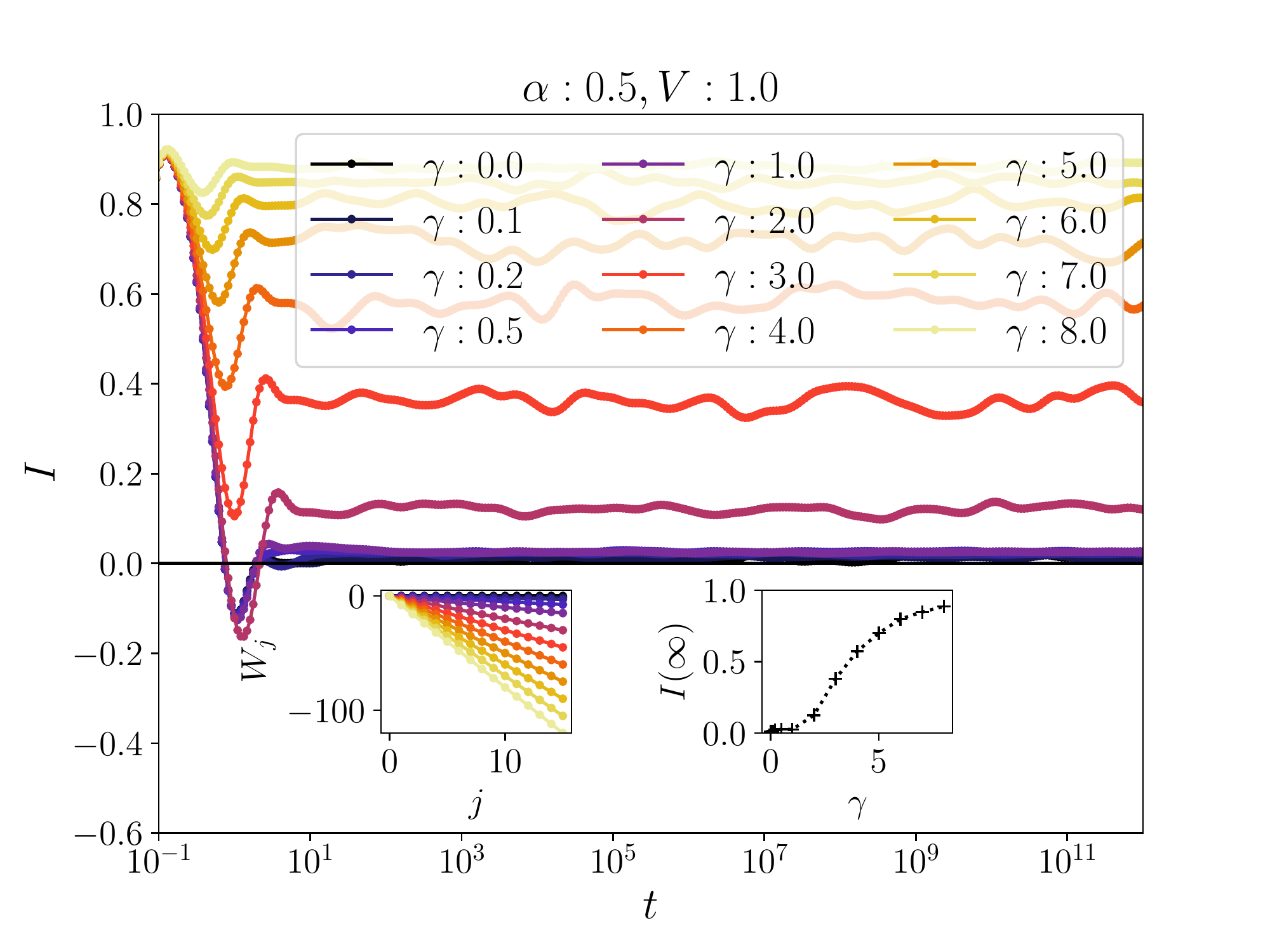}
	\caption{The $L = 16$ study of a hypothetical imbalance experiment, where an initial charge-density wave relaxes under unitary time-evolution. The different curves represent different field strengths $\gamma$ with $\alpha = 0.5$. Inset left: Visualization of the corresponding potential. Inset right: Average imbalance $\mathcal{I}$ at times $10^{13}\geq t/J \geq 10^{11}$. The numerical curves have been smoothed by convolution with a Gaussian, $w(n) = e^{-(n/\sigma)^2/2}$, with $\sigma = 4$.}
	\label{fig:imbalance}
\end{figure}

In Fig.~\ref{fig:imbalance}, we show exact diagonalization results of the time-evolution of an initial CDW state under unitary time-evolution of \eqref{equ:hamiltonian}.
After initial transient behavior, the imbalance \eqref{equ:imbalance} remains non-zero for arbitrarily long times for sufficiently large fields.

\paragraph{Discussion.}
In summary, we have shown that the experimentally natural case of a finite system in an electric field, a simple Wannier-Stark-localized system, shows properties that coincide with those of the MBL phase.
While the case of a purely uniform field, $\alpha = 0$, remarkably turns out to be a non-generic limit, even moderate curvature gives consistent and robust MBL-like features.
The bipartite entanglement entropy $S(t)$ exhibits a slow, logarithmic growth to a value much larger than that obtained in the non-interacting case.
The spectral statistics, a dynamics-independent measure for localization, are Poissonian.
And finally, localization seems equally persistent in a now standard imbalance experiment, where the relaxation of CDW order is measured.

In the limit of large system sizes, the energy density can grow without bound on account of the unbounded potential implied by a uniform compound of the electric field. This precludes an asymptotic definition of exponential localization, which should however not pose a problem in practice given the attainability of short localization lengths. While the  set-up of Stark-MBL is quite different from conventional disorder MBL, the similar phenomenology is quite striking.  The MBL phase is described in terms of $l$-bits, while other disorder-free localized systems have also identified integrals of motion. Here, $l$-bits emerge naturally and robustly as Stark locators, again without the use of any quenched disorder. While finalizing this manuscript, we became aware of the related work by van Nieuwenburg \textit{et al.}~\cite{vanNieuwenburg2018}.

\paragraph{Acknowledgments.} MS acknowledges financial support from the CM-CDT under EPSRC (UK) grant number EP/L015110/1.  CAH acknowledges financial support from the TOPNES programme under EPSRC (UK) grant number EP/I031014/1.  This research was supported in part by the National Science Foundation under Grant No.\ NSF PHY-1125915.  Part of this work was performed at the Aspen Center for Physics, which is supported by National Science Foundation grant PHY-1066293. FP acknowledges the support of the DFG Research Unit FOR 1807 through grants no.\ PO 1370/2-1, TRR80, the Nanosystems Initiative Munich (NIM) by the German Excellence Initiative, and the European Research Council (ERC) under the European Union’s Horizon 2020 research and innovation program (grant agreement no.\ 771537).

\bibliographystyle{apsrev4-1}
\bibliography{2018_stark}

\section{Supplemental Material}
 
  \section{Semi-analytic calculation of entanglement entropy}

 \subsection{Model and initial condition}
 In the following we develop an approximate analytical description of the entanglement entropy growth in Stark-MBL from a specific product-state initial condition, which allows us to circumvent full diagonalization of the Hamiltonian, Eq.~(1) of the main text. We shall assume that the number of sites, $L$, is even, and that the entanglement entropy is calculated across a cut between sites $(L/2)-1$ and $L/2$.  We choose our zero of potential energy to lie in the middle of that bond.  We shall often ignore the $V$-term, assuming that it is too weak to modify the form of the many-body eigenfunctions significantly, i.e. we consider the leading effect of V on energies, not wave functions.
 
 Our initial condition involves placing fermions at the sites $S = \{ j_k \}$, where $k=1,2,\ldots,M$, and $M$ is the total number of fermions.  Let $M = M_L + M_R$, where $M_L$ and $M_R$ are respectively the number of fermions to the left and right of the cut, and let us assume that the list $S$ is in order from left to right (i.e.\ that $j_k < j_{k+1}$ for all relevant $k$).  This initial condition can be written in second-quantized form as
 \begin{equation}
 \label{state}
 \ket{\psi} = c^\dagger_{j_M} c^\dagger_{j_{M-1}} \ldots c^\dagger_{j_2} c^\dagger_{j_1} \ket{0}, 
 \end{equation}
 where $\ket{0}$ is the fermionic vacuum. Let us note that the electric potential energy of this state is given by
 \begin{equation}
 E_0 = \gamma \sum_{k=1}^M  j_k,
 \end{equation}
 where we have neglected the contribution of $\alpha$.
 \subsection{Writing the initial condition in the many-body eigenbasis}
 Clearly we can only do this analytically if we ignore $V$, so we shall do that for now.  In that case, we just need to know how to write the on-site creation operator in terms of creation operators for the Stark-localized states.  If we assume the strong-field limit, i.e.\ that $\gamma \gg J$, then we may suppose that
 \begin{equation}
 c^\dagger_j = \sqrt{1 - 2 \left( \frac{J}{\gamma} \right)^2} \, \beta^\dagger_j + \frac{J}{\gamma} \left( \beta^\dagger_{j+1} + \beta^\dagger_{j-1} \right), \label{trans}
 \end{equation}
 where $\beta^\dagger_j$ creates a fermion in the single-particle Stark-localized state centered at position $j$.  This assumes that the Stark-localized states are restricted to three sites (the central, or `locator', site and its two nearest neighbors).  It also does not account for the mutual orthogonality of the different Stark-localized states, and it does not work at the very ends of the chain.
 
 Substituting (\ref{trans}) into (\ref{state}), we obtain the following:
 \begin{equation}
 \ket{\psi} = \sum_Q f_Q \beta^\dagger_{Q(j_{M})} \beta^\dagger_{Q(j_{M-1})} \ldots \beta^\dagger_{Q(j_{2})} \beta^\dagger_{Q(j_{1})} \ket{0}. \label{stateQ}
 \end{equation}
 Here $Q$ is an operation that, for each original site label, either moves it one place to the left, moves it one place to the right, or leaves it alone.  The factor $f_Q$ is given approximately by
 \begin{equation}
 f_Q \approx \left( \frac{J}{\gamma} \right)^{\lambda_Q + \rho_Q}, \label{stateweight}
 \end{equation}
 where $\lambda_Q$ counts the number of original site labels that are shifted to the left by $Q$, and $\rho_Q$ the number that are shifted to the right.
 \subsection{Calculating the many-body energies of the components of the state}
 In order to time-evolve (\ref{stateQ}), we attach to each many-body state in the sum a factor  of $e^{-i E_Q t}$, where $E_Q$ is the state's energy.  As we know from Ref.~\cite{Serbyn2013a}, we need to be more precise about the tails of the single-particle wave functions at this stage if we want to capture the entanglement evolution at late times.
 
 The energy $E_Q$ is composed of two parts:\ the electric potential energy, and the repulsive interaction energy of each pair of its constituent particles.  The electric potential energy can be calculated by noting which particles have moved compared to the reference state:
 \begin{equation}
 E_{Q,{\rm elec}} = E_0 + \gamma \left( \rho_Q - \lambda_Q \right).
 \end{equation}
 The repulsive interaction energy may be roughly calculated by taking the probability density in the tail of one Stark-localized state at the center of the other and multiplying it by  $V$.  For two Stark-localized states with locators separated by $\xi$ lattice sites, this gives an energy of approximately
 \begin{equation}
 V \left( \frac{J}{\gamma} \right)^{2\xi -2}.
 \end{equation}
 Thus the complete energy of the state-component $Q$ is given approximately by
 \begin{equation}
 E_Q \approx E_0 + \gamma \left( \rho_Q - \lambda_Q \right) + \frac{V}{2} \sum_k \sum_{p \ne k} \left( \frac{J}{\gamma} \right)^{2 \left\vert Q(j_k) - Q(j_p) \right\vert -2}, \label{manyen}
 \end{equation}
 where the sums over $k$ and $p$ run from $1$ to $M$.
 \subsection{Calculating the entanglement entropy as a function of time}
 We can thus approximate the state of the $M$-fermion system at an arbitrary time $t$ as
 \begin{equation}
 \ket{\psi(t)} = \sum_Q f_Q e^{-i E_Q t} \beta^\dagger_{Q(j_{M})} \beta^\dagger_{Q(j_{M-1})} \ldots \beta^\dagger_{Q(j_{2})} \beta^\dagger_{Q(j_{1})} \ket{0}, \label{arbtime}
 \end{equation}
 where $f_Q$ is given by (\ref{stateweight}) and $E_Q$ is given by (\ref{manyen}).  To turn this into a density operator it is more convenient to write it in first-quantized notation:
 \begin{eqnarray}
 \ket{\psi(t)} = \sum_Q f_Q e^{-i E_Q t} \Big[ \ket{Q(j_1)} \otimes \ket{Q(j_2)} \otimes \ldots \\ \otimes \ket{Q(j_{M-1})} \otimes \ket{Q(j_M)} \Big],
 \end{eqnarray}
 where $\ket{j}$ is the single-particle state in which the particle is on site $j$. From this we obtain the density operator. We trace out the states on the right-hand side of the cut to construct the reduced density matrix on the left, and then use that to calculate the entanglement entropy $S(t)$. The results can be see in Fig.~2 of the main text.
 
\section{Level statistics in the case of too little field non-uniformity}

In this section we present the case in which the field non-uniformity is not strong enough, hence in which case $\alpha$ is too small.
We recall from the main text that the model reads
\begin{multline}
\label{equ:hamiltonian}
\hat{H} =  \frac{J}{2} \,\sum_{j=0}^{L-2} \left( c^\dagger_j c^{\phantom{\dagger}}_{j+1} + \text{h.c.} \right)  + \sum_{j=0}^{L-1} W_j \left(n_j -\frac{1}{2}\right) \\ + V \sum_{j=0}^{L-2} \left(n_j -\frac{1}{2}\right) \left(n_{j+1} -\frac{1}{2}\right)  . 
\end{multline}
Here the operator $c^{\dagger}_{j}$ creates a fermion on lattice site $j$, and the associated number operator $n_j = c^{\dagger}_{j} c_{j}$. $J/2$ is the hopping matrix element between neighboring sites, $V$ is the strength of the nearest-neighbor repulsion, and $L$ is the number of sites of the lattice, the boundary conditions of which we take to be open.

$W_j$ is the on-site potential energy due to the applied electric field.  For a uniform field $\gamma$, it takes the form $W_j = -\gamma j$.  For a non-uniform field, it acquires some curvature, $W_j = - \gamma j + \alpha j^2/L^2$.   This means that the potential has the same value and slope as the linear one at $j=0$, but differs at the other end of the chain by $\alpha$ from the purely linear case.

\begin{figure}
	\includegraphics[width=\linewidth]{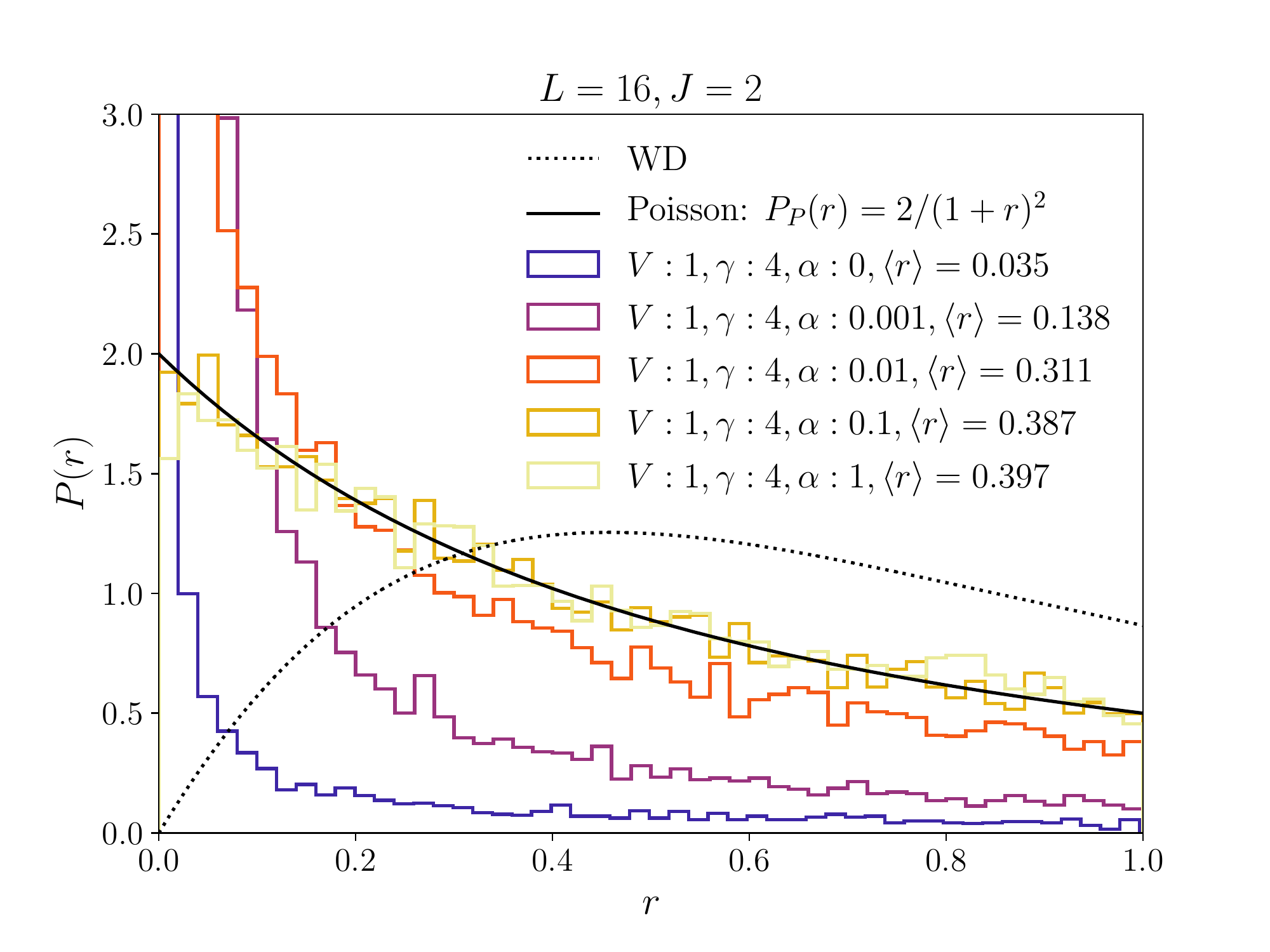}
	\caption{Illustration of the many-body energy level statistics for the half-filled sector as shown in Fig.~3 of the main text. Again, we study the probability distribution of the gap-ratio parameter, $r_n = \min(\delta_n/\delta_{n+1}, \delta_{n+1}/\delta_n)$ (where $\delta_n$ is the gap between the $n^{\text{th}}$ and $(n-1)^{\text{th}}$ energy eigenvalues). Here we illustrate the case of small or zero $\alpha$ in which Stark MBL is absent, which we attribute to the many exact degeneracies in the non-interacting problem. This creates a disproportionately large weight for small $r_n$ in the many body spectrum clearly illustrated in the case of small or zero $\alpha$.}
		\label{supp:level_fail}
\end{figure}

Numerically diagonalizing (\ref{equ:hamiltonian}), and then obtaining the level-statistics and more specifically the gap-ratio parameter, $r_n = \min(\delta_n/\delta_{n+1}, \delta_{n+1}/\delta_n)$ (where $\delta_n$ is the gap between the $n^{\text{th}}$ and $(n-1)^{\text{th}}$ energy eigenvalues) allows us to make predictions about the existence of a localized phase. In Fig.~\ref{supp:level_fail} we show the case for small and zero $\alpha$ in which the system exhibits neither Poisson nor Wigner-Dyson level statistics. The disproportionately large weight for small $r_n$ in this case is in line with our reasoning that this failure to exhibit Poisson or Wigner-Dyson level statistics is due to the many exact degeneracies in the non-interacting problem.
\section{The cross-over from Wigner-Dyson to Poisson}
\begin{figure}
	\includegraphics[width=\columnwidth]{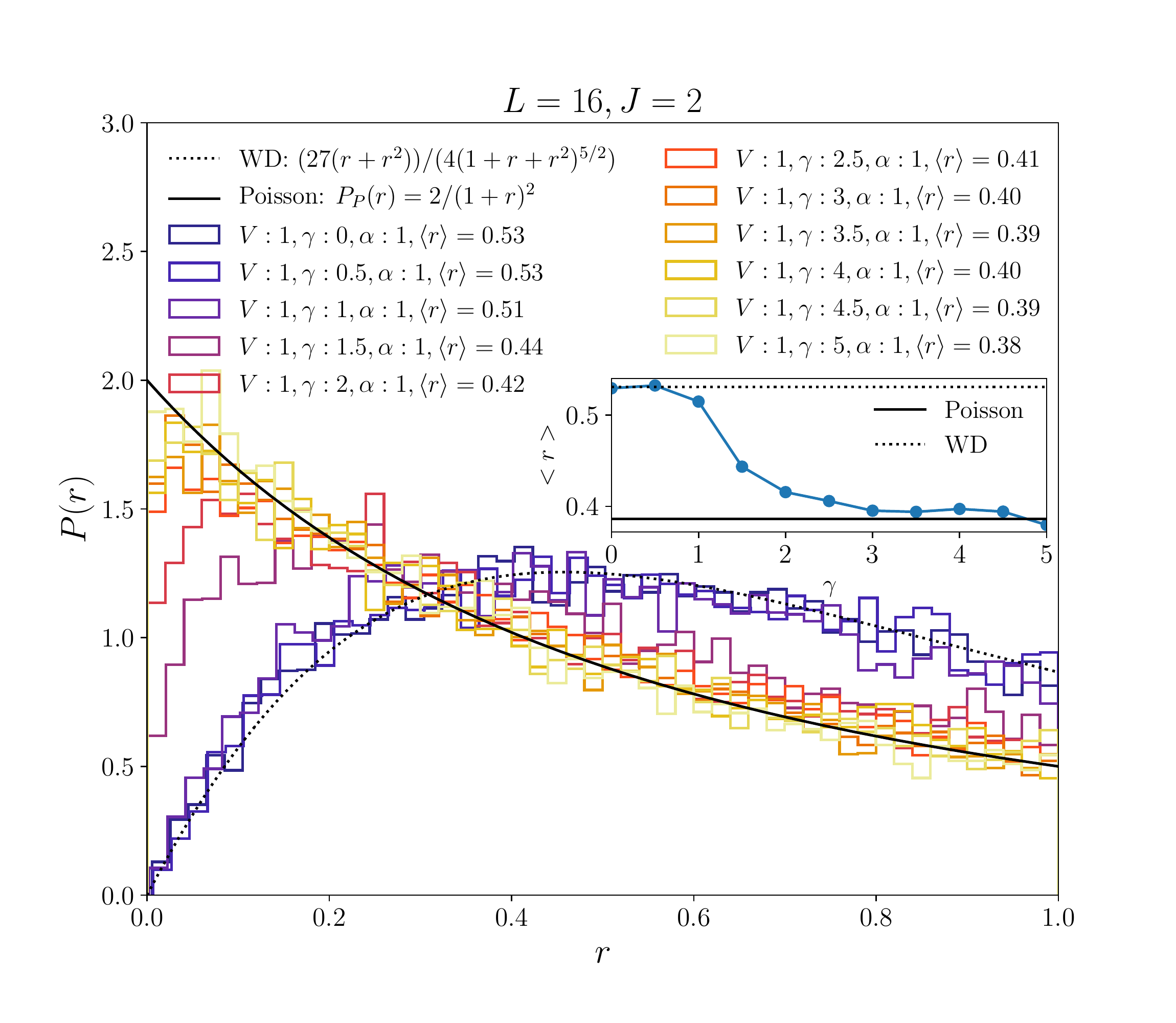}
	\caption{Illustration of the many-body energy level statistics for the half-filled sector as shown in Fig.~3 of the main text. Again, we study the probability distribution of the gap-ratio parameter, $r_n = \min(\delta_n/\delta_{n+1}, \delta_{n+1}/\delta_n)$ (where $\delta_n$ is the gap between the $n^{\text{th}}$ and $(n-1)^{\text{th}}$ energy eigenvalues). Here we illustrate the cross-over from Wigner-Dyson to Poisson level statistics for fixed $\alpha$ but changing uniform field strength $\gamma$. The crossover is clearly visible with the inset showing the average gap-ratio parameter $\langle r \rangle$ as a function of $\gamma$.}
	\label{supp:WDtoPoiss}
\end{figure}
\begin{figure}
	\includegraphics[width=\columnwidth]{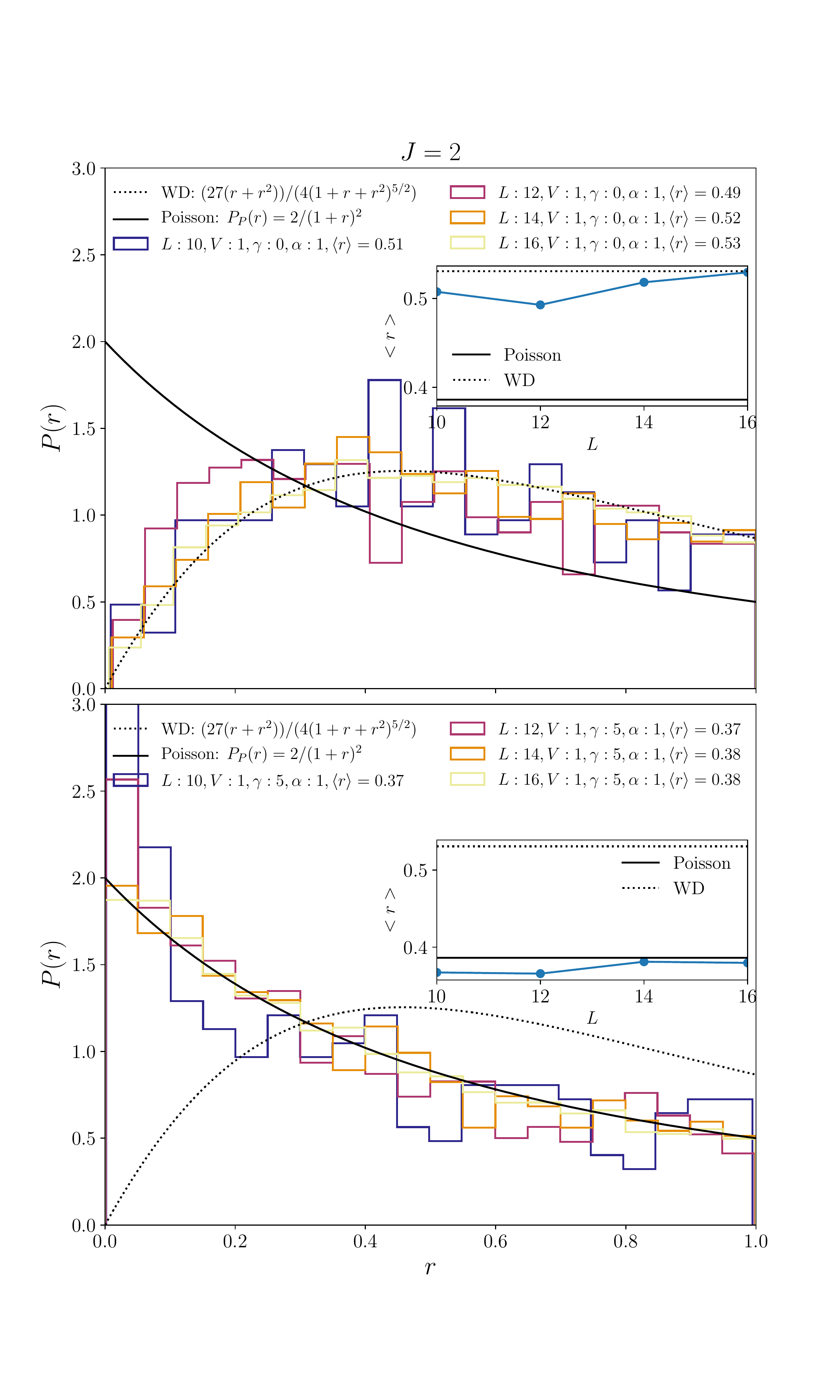}
	\caption{Finite-size scaling of the gap-ratio parameter as described in previous figures. The two panels show the distribution of the gap-ratio parameter for two different values of the field-strength $\gamma$, namely $\gamma = 0$ and $\gamma = 5$, for different system sizes $L$. As far as the presented system sizes are concerned, we do not observe any significant flow towards the Poisson or Wigner-Dyson limit as the system size $L$ grows and therefore we expect the presented physics to be observable in any realistic finite experimental setup. The insets shows the average gap-ratio parameter as a function of system size $L$.}
	\label{supp:finitesize}
\end{figure}
In this section we present additional results on the cross-over from Wigner-Dyson (WD) to Poisson statistics in the case of vanishing field $\gamma$ with fixed $\alpha$. In this \textit{finite} lattice setup this means that as $\gamma \rightarrow 0$, the localization length of the single particles will grow larger than the length of the chain, thus making the system effectively delocalized. We note that this relies on $\alpha$ being chosen small enough such that it does not provide any effective single-particle localization itself. The level statistics in this case were obtained as described in the previous section. The results are shown in Fig.~\ref{supp:WDtoPoiss}. We clearly see the cross-over from WD to Poisson statistics for increasing field strength. The inset illustrates this particularly well, displaying the average gap-ratio parameter $\langle r \rangle$ as a function of $\gamma$. As a reminder we note that the expected value for Poisson statistics is $\langle r \rangle_{\text{P}} = 0.39$, while for WD statistics, $\langle r \rangle_{\text{GOE}} = 0.53$ (in the symmetry class of the Gaussian Orthogonal Ensemble) is the expected value. 

We may further investigate the finite-size scaling of the gap-ratio parameter $r$. The results of this are shown in Fig.~\ref{supp:finitesize}. As can be seen from the finite-size scaling of small systems in Fig.~\ref{supp:finitesize}, there is no significant flow towards either the Poisson or the WD limit when the system is made larger. As such we expect the phenomenology of Stark MBL, including the cross-over to WD statistics to be visible in realistic finite size experimental setups. It remains an interesting open question how the gap-ratio parameter would behave in the limit of very large system size.

\end{document}